# Preface: Long-range Interactions and Synchronization

Shamik Gupta, Ramakrishna Mission Vivekananda University, India

Romain Bachelard, Universidade Federal de São Carlos, Brazil

Tarcísio Marciano da Rocha Filho, Universidade de Brasília, Brazil

Spontaneous synchronization is a general phenomenon in which a large population of coupled oscillators of diverse natural frequencies self-organize to operate in unison. The phenomenon occurs in physical and biological systems over a wide range of spatial and temporal scales, e.g., in electrochemical and electronic oscillators, Josephson junctions, laser arrays, animal flocking, pedestrians on footbridges, audience clapping, etc. Besides the obvious necessity of the synchronous firings of cardiac cells to keep the heart beating, synchrony is desired in many man-made systems such as parallel computing, electrical power-grids. On the contrary, synchrony could also be hazardous, e.g., in neurons, leading to impaired brain function in Parkinson's disease and epilepsy. Due to this wide range of applications, collective synchrony in networks of oscillators has attracted the attention of physicists, applied mathematicians and researchers from many other fields. An essential aspect of synchronizing systems is that long-range order naturally appear in these systems, which questions the fact whether long-range interactions may be particular suitable to synchronization. In this context, it is interesting to remind that long-range interacting system required several adaptations from statistical mechanics à la Gibbs Boltzmann, in order to deal with the peculiarities of these systems: negative specific heat, breaking of ergodicity or lack of extensivity. As for synchrony, it is still lacking a theoretical framework to use the tools from statistical mechanics. The present issue presents a collection of exciting recent theoretical developments in the field of synchronization and long-range interactions, in order to highlight the mutual progresses of these twin areas.

- The paper [1] "**Phase diagram of noisy systems of coupled oscillators with a bimodal frequency distribution**" by Campa studies systems of globally-coupled phase oscillators in presence of Gaussian, white noise, for the case when the distribution of natural frequencies of the oscillators is bimodal and its analytical continuation in the complex plane has only a few poles in the lower half plane. The stationary-state phase diagram of the oscillator system is shown to exhibit a very rich behaviour, with regions characterized by synchronized states, periodic states, and bistability. With the increase in noise strength, the author argues that the phase diagram tends to coincide with the one for unimodal frequency distributions.

- The paper [2] "**Traveling chimera states**" by Omelchenko studies existence of traveling chimera or broken-symmetry states in a prototype model involving a heterogeneous network of nonlocally coupled phase oscillators. The study invokes a long-time coarse-grained dynamics given by the so-called Ott-Antonsen equation. It is demonstrated that the equation provides a reliable description of traveling chimera states for heterogeneous oscillators with Lorentzian distribution of natural frequencies, while it fails to do so for the case of identical oscillators.

- In the paper [3] "**Linear response theory for coupled phase oscillators with general coupling functions**," Terada *et al.* develop a linear response theory for a wide class of globally-coupled phase oscillator systems prepared in a nonsynchronized reference state and subject to an external force. This allows them to compute the asymptotic value of the order parameter in the system. Remarkably, the proposed theory applies for any natural frequency distribution of the oscillators, as well as for any phase-lag parameter and any value of the time-delay parameter characterizing the interaction among the oscillators.

- In the work [4] "**Synchronization in disordered superconducting arrays**," Wiesenfeld studies the effect of quenched disorder on the synchronization properties of a class of circuits known as Josephson junction arrays. In particular, a comparative assessment of four different dynamical scenarios involving Josephson junction arrays is offered through a mapping of the corresponding dynamics onto a tractable model of globally-coupled phase oscillators.

- In the paper [5] "**The Winfree model with heterogeneous phase-response curves: analytical results**," Pazó *et al.* study the Winfree model of globally-coupled phase oscillators, considered an important milestone in the mathematical description of collective synchronization. In particular, they consider an extension of the model in which the oscillator natural frequencies are taken to be heterogeneous, as also the phase-response curves determining the response of an oscillator to an input pulse. For this model, an approximate dynamics is derived in which the oscillators are coupled through their phase differences, and for which the results are found to be fully consistent with those for the original model for Lorentzian heterogeneities.

- The paper [6] "**Synchronized states of one-dimensional long-range systems induced by inelastic collisions**" by Joyce *et al.* reports on a family of simple one dimensional particle systems with long-range interactions and subjected to inelastic interactions. The work unveils that the system evolves towards states that are highly ordered in phase space, with particle motions that are periodic and synchronized in relative phase. An interesting result obtained for a self-gravitating system is that these states show a remarkable stability when the inelastic interactions are turned off, with the phase space order persisting much longer than expected from a long-range system relaxing toward equilibrium.

- In the paper [7] "**Synchronization in complex networks with long-range interactions**" Rakshit *et al.* investigate the dynamics of a complex network that includes all possible k-path couplings between all pairs of nodes in the network. The work is devoted to the derivation of local and global asymptotic stability conditions for complete synchronization in the system.

- The paper [8] "**Ageing in mixed populations of Stuart–Landau oscillators: the role of diversity**" by Sahoo *et al.* studies the question of ageing transition in a population of globally coupled Stuart–Landau oscillators, namely, the increase in time of the number of inactive (or non-oscillatory) units. The system has an initial population that is either a mixture of active and inactive oscillators or an ensemble of active oscillators with a mixture of distinct frequencies. It is shown that the ageing transition is determined by the degree of dynamical diversity in the system.

- In [9] "**Criticality of spin systems with weak long-range interactions**", Defenu *et al.* investigate systems that lie at the transition between short- and long-range, displaying on the one hand the extensivity used in Bolzmann-Gibbs approach, and on the other hand a universality class that departs from the true short-range systems. These "weak long-range" systems are reviewed using both classical and quantum spin systems. The use of the functional renormalization group allows the authors to compute the critical exponents of the system, and to characterize in detail the signatures in criticality of the transition from short- to long-range interactions.

- In [10] "**One-dimensional quantum many body systems with long-range interactions**", Maity and coauthors review the dynamics of quantum one-dimensional chains, focusing on the effect of long-range interactions on the growth of the entanglement entropy, the propagation of mutual information and non-equilibrium phase transitions. They study in more detail the long-range version of the Kitaev chain, identifying the range of interaction from which long-range features arise. Motivated by the relevance of these effects for superconductor theory, they pay particular attention to the role of pairing terms.

- In [11] "**Lieb–Robinson bounds for open quantum systems with long-ranged interactions**", Sweke and coauthors derive Lieb-Robinson bounds for open quantum systems with power-law interactions. These bounds describe the propagation of quantum correlations, entanglement and quantum information, so they may be of great relevance for quantum simulators based on arrays of Rydberg atoms or trapped ions. The locality bounds derived present a power-law decay, which is inherited from the considered interaction.

- In [12] "**Equilibrium time-correlation functions of the long-range interacting Fermi–Pasta–Ulam model**", Di Cintio *et al.* investigate the Fermi–Pasta–Ulam chain with power-law interactions, showing the presence of long-wavelength propagating modes. These numerical findings are in excellent agreement with predictions from a linear-dispersion theory. The simulations also support the validity of the dynamical scaling hypothesis. A surprising ballistic behavior is reported for $1/r^2$ interactions, showing that the celebrated FPU model and its variants still hold many surprises.

- In [13] "**Fluctuation-dominated phase ordering at a mixed order transition**", Barma and coauthors discuss the order parameter correlation function in presence of a mixed order transition. Characterized by a diverging correlation length and a discontinuity of the order parameter, these transitions are shown to exhibit fluctuation-dominated order for an Ising model with long-range interactions. It is suggested that this feature may be observed in a variety of mixed-order transition systems.

- In [14] "**Long-range interaction induced collective dynamical behaviors**", Sathiyadevi *et al.* study the interplay between repulsive and attractive coupling in a collection of Stuart-Landau limit cycle oscillators. They report on the swing of the synchronized states, since their stability is challenged by the emergence of solitary or cluster states. The authors are able to derive some bounds which predict accurately the transition from one regime to the other, shedding light on the synchronization of systems with competing interactions.

- In [15] "**Prethermalization in the transverse-field Ising chain with long-range interactions**", Mori studies the out-of-equilibrium dynamics of a long-range Ising chain with power-law interactions. The prethermalization is characterized by a separation in timescales, one being associated with the quasi-conserved quantities, the other stemming from the growth of quantum fluctuations. The discrete truncated Wigner approximation is shown to describe properly this initial dynamics, while the tools necessary to study the full equilibration of a large quantum system remain to be identified.

- In[16] "**Entropy production and Vlasov equation for self-gravitating systems**", Farias and coauthors discuss the validity of the Vlasov equation to describe the violent relaxation process that characterizes the initial transition to long-lived out-of-equilibrium states in long-range systems. Focusing on self-gravitating systems, the authors show that the growth of entropy production is in excellent agreement with full molecular dynamics. Their work highlights the importance of dealing appropriately with the coarse-grained phase-space to describe the loss of resolution and the consequent entropy production.

- In [17] "**Periodically driven integrable systems with long-range pair potentials**", Nandy *et al.* investigate the periodic driving of a Kitaev chain with power-law interactions. The authors identify a critical range above which the system displays particularly long equilibration times, as measured from a reduced density matrix. This behavior is also true for correlation functions, and mutual information also exhibits an interesting transition, with a multiple-light-cone structure. The work opens new avenues for the study of the Floquet dynamics of integrable systems.